# An old efficient approach to anomalous Brownian motion

V. Lisy and J. Tothova

Department of Physics, Technical University of Kosice, Park Komenskeho 2,
04200 Kosice, Slovakia

**Abstract.** A number of random processes in various fields of science is described by phenomenological equations containing a stochastic force, the best known example being the Langevin equation (LE) for the Brownian motion (BM) of particles. Long ago Vladimirsky (1942) proposed a simple method for solving such equations. The method, based on the classical Gibbs statistics, consists in converting the stochastic LE into a deterministic one, and is applicable to linear equations with any kind of memory. When the memory effects are taken into account in the description of the BM, the mean square displacement of the particle at long times can exhibit an "anomalous" (different from that in the Einstein theory) time dependence. In the present paper we show how some general properties of such anomalous BM can be easily derived using the Vladimirsky approach. The method can be effectively used in solving many of the problems currently considered in the literature. We apply it to the description of the BM when the memory kernel in the Volterra-type integro-differential LE exponentially decreases with the time. The problem of the hydrodynamic BM of a charged particle in an external magnetic field is also solved.

## I. INTRODUCTION

The elucidation of phenomena that contradict to the traditional views often leads to a deeper understanding of the existing models or even to the creation of new theories and conceptions. Such "anomalous" effects always attract a lot of attention of scientists. During the past decades, a continuously increasing interest is observed to the so called anomalous diffusion in various fields of science and technology [1]. It has been established that a vast variety of physical, chemical, biological and other natural phenomena can be adequately described by random processes exhibiting at long times a behavior, which is characterized by a nonlinear dependence of the variance of these processes on time. If we remember the famous Einstein formula [2] for the long-time behavior of mean square displacement (MSD) of the Brownian particle (BP), $\langle[\mathbf{r}(t) - \mathbf{r}(0)]^2\rangle \sim t$, then in the case of anomalous diffusion the MSD dependence on the time is $\sim t^\alpha$. We speak about sub-diffusion if $\alpha < 1$ and super-diffusion for $\alpha > 1$. Mathematically, the anomalous diffusion is described by fractional differential equations or Volterra-type integral equations with a colored stochastic force driving the particle [3, 4]. These equations generalize the well-known Langevin equation (LE) [5] designed for the description of memory-less Brownian motion (BM). In the present contribution we propose a very simple way of solution of the generalized LE (GLE). The method was published long ago by Vladimirsky [6] but, to our knowledge, so far it has not been applied for solving the GLE. Using this method, we first show how some general properties of anomalous diffusion can be derived. Then we consider a special kind of the memory in the system associated with the driving stochastic force represented by an Ornstein-Uhlenbeck process. An exact solution of the GLE is given for this case. The Vladimirsky method can be used for the description of the BM in linear classical systems with any kind of memory. We illustrate its effectiveness on the example of the hydrodynamic BM of a charged particle moving in an incompressible liquid in an external magnetic field.

## II. EFFICIENT RULE FOR SOLVING THE LANGEVIN EQUATION

In Ref. [6], Vladimirsky considered a method for the evaluation of the averaged products of two thermally fluctuating quantities related to different moments of time. The difficult task of the calculation of such means using the Gibbs method was reduced to a much simpler solution of equations of motion. The obtained rules of finding the time correlation functions are thus consistent with the general principles of statistical mechanics and simultaneously with the linear phenomenological equations of motion for the studied quantities. For the purposes of the present paper, let us consider the LE for the coordinate $x$ of the BP. If we are interested in finding the MSD of the particle, $\xi(t) = \langle \Delta x^2(t) \rangle = \langle [x(t) - x(0)]^2 \rangle$, according to the Vladimirsky rule we have merely to substitute $x(t)$ in the LE by $\xi(t)$ and replace the stochastic force driving the particle with the constant "force" $f = 2k_B T$, where $k_B$ is the Boltzmann constant and $T$ the temperature. Of course, having the dimension of energy, $f$ is not a true force; it only plays this role in the equation for the particle "position" $\xi(t)$. This fictitious force begins to act on the particle at the time $t = 0$. Up to this moment the particle is nonmoving so that the equation of motion must be solved with the initial conditions $\xi(0) = V(0) = 0$, where we have introduced the "velocity" $V(t) = d\xi(t)/dt$. It is important that at the initial moment the whole studied system is assumed to be in the thermodynamically equilibrium state. From the point of view of the macroscopic equations of motion such initial state corresponds to the long-lasting rest of the system. Thus the values of all the parameters entering the equations of motion are specified for any $t \leq 0$. This allows one to study not only equations for $x(t)$ in the form of differential equations (that require only the specifications of the initial values of the quantities) but also any Volterra-type integral equations. Such equations can describe systems for which the whole history of its motion is important. The discussed method is not limited to a concrete law of the aftereffect. Its advantage is that it is applicable to any kind of the memory in the studied (linear) system and the only requirement is the linearity and existence of the unique solution at the given initial conditions.

Let us illustrate the described rule by a simple example of a harmonic oscillator under the influence of the thermal Langevin force $f(t)$, described by the equation of motion for the position $x(t)$, $m\ddot{x} + \gamma \dot{x} + m\omega_0^2 x = f(t)$ (the parameters $m$, $\gamma$ and $\omega$ can be regarded as the mass of a particle, the friction constant and the oscillator frequency, respectively). After multiplying this equation by $x(t)$, then using the identity $x\ddot{x} = d(x\dot{x})/dt - \dot{x}^2$, the averaging assuming the statistical independence of $x$, $\dot{x}$ and $f$, the original equation can be with the help of the equipartition theorem $m\langle \dot{x}^2 \rangle = k_B T$ rewritten as $md^2\langle x^2 \rangle/dt^2 + \gamma d\langle x^2 \rangle/dt + 2m\omega_0^2 \langle x^2 \rangle = 2k_B T$. If we now subtract the equation obtained for the correlator $\langle x(t)x(0) \rangle$, we find $m\ddot{\xi}(t) + \gamma \dot{\xi}(t) + m\omega_0^2 \xi(t) + m\omega_0^2(\langle x^2(t) \rangle - \langle x^2(0) \rangle) = 2k_B T$. For stationary random processes the last term on the left hand side is equal to zero. We thus come to the final equation $m\ddot{\xi} + \gamma \dot{\xi} + m\omega_0^2 \xi = 2k_B T$ and the evident initial conditions $\xi(0) = \dot{\xi}(0) = 0$. Analogously more complicated phenomenological (but linear) stochastic equations can be converted to deterministic ones, e.g. in the case when the friction force $\gamma \dot{x}$ s replaced by a convolution of the time dependent memory kernel $\Gamma(t)$ and the particle velocity $\dot{x}(t)$. This corresponds to the GLE considered in the next section.

## III. SOLVING THE GENERALIZED LANGEVIN EQUATION

So, already in 1942 [6] an effective approach to various problems that are now extensively studied in the literature was available. Let us consider a few applications of this approach to the problems of anomalous diffusion. Many non-Markovian processes are described by the GLE [3, 4]

$$M\dot{v} + \int_0^t \Gamma(t-t')v(t')\,\mathrm{d}t' = \eta(t), \tag{1}$$

where, for the anomalous BM, $M$ could be the mass of the BP with the velocity $v(t) = \dot{x}(t)$. The memory in the system is described by the kernel $\Gamma(t)$, and $\eta(t)$ is a stochastic force with zero mean. The fluctuation-dissipation relation dictates that the condition $\langle \eta(0)\eta(t)\rangle = k_B T \Gamma(t)$, $t > 0$, must be satisfied [3, 7]. One is usually interested in finding the MSD of the particle, $\xi(t) = \int_0^t V(\tau)\,\mathrm{d}\tau$, its time-dependent diffusion coefficient $D(t) = V(t)/2$, or the velocity autocorrelation function (VAF) $\Phi(t) = \dot{D}(t)$ [8]. The initial conditions for $\xi(t)$ and $V(t)$ have been discussed above. Obviously, also the condition $\dot{V}(0) = 2k_B T / M$ must hold. It follows from the equation that concretizes the Vladimirsky rule for this special case,

$$M\dot{V} + \int_0^t \Gamma(t-t')V(t')\,\mathrm{d}t' = 2k_B T. \tag{2}$$

Equation (2) can be easily solved for the specific memory kernels considered in the literature. It also allows one to derive the properties of this kind of anomalous diffusion that have been obtained so far by a more complicated manner. For example, taking the Laplace transformation $\mathcal{L}$ of Eq. (2) and using the initial conditions, we obtain for $\tilde{V}(s) = \mathcal{L}\{V(t)\}$ and $\tilde{\xi}(s) = \mathcal{L}\{\xi(t)\}$

$$\tilde{V}(s) = \frac{2k_B T}{s} \frac{1}{Ms + \tilde{\Gamma}(s)}, \quad \tilde{\xi}(s) = \frac{\tilde{V}(s)}{s}. \tag{3}$$

It immediately follows from this equation that if at small $s$ the kernel $\tilde{\Gamma}(s)$ behaves as $\tilde{\Gamma} \propto s^\nu$ and the MSD at long times is $\xi(t) \propto t^\alpha$, then there are two possibilities: i) if $\nu < 1$, then $\alpha = \nu + 1$ [9], and the VAF $\Phi(t)$ decays to zero as $t \to \infty$ (this corresponds to the irreversibility condition and, consequently, the ergodicity of the variable $v$ [10]), ii) if $\nu \geq 1$, then $\alpha = 2$ (ballistic motion) and $\Phi(t)$ converges to a constant as $t \to \infty$. We have sub-diffusion ($\alpha < 1$) if $\nu < 0$ and super-diffusion ($\alpha > 1$) if $\nu > 0$. The Einstein diffusion corresponds to $\nu = 0$. At short times the result is consistent with the equipartition theorem,

$$D(t) \approx \frac{k_B T}{M} t, \quad \xi(t) \approx \frac{k_B T}{M} t^2, \, t \to 0, \tag{4}$$

independently of a concrete form of $\Gamma(t)$.

Note that if $\Gamma(t) \propto t^{-\alpha}$, $0 < \alpha < 1$, the second term on the left side of Eq. (1) can be expressed through the fractional derivative $_0 D_t^{\alpha-1} v(t)$. Equation (2) can be thus used also for effective solving of fractional LE for the BM [11].

Now let us consider an example, in which the memory kernel is specified. We chose $\Gamma(t)$ in the Ornstein-Uhlenbeck form $\Gamma(t) = (\gamma^2/m)\exp(-\gamma t/m)$. Then the force $\eta(t) = m\dot{u}(t)$ corresponds to the solution of the usual LE $m\dot{u}(t) + \gamma u(t) = f(t)$. Here $u(t)$ can be the velocity of a BP, $\gamma$ is a constant friction coefficient, and $f(t)$ the stochastic (white noise) force. The motion of the BPs with the mass $M$ can be thus interpreted as being induced by the force $\eta$ caused by the particles of mass $m$.

The Laplace transformation of Eq. (2) yields

$$\tilde{V}(s) = \frac{2k_B T}{Ms}\left(s + \frac{\gamma}{m}\right)\left(s^2 + \frac{\gamma}{m}s + \frac{\gamma^2}{mM}\right)^{-1} \tag{5}$$

The denominator in this expression can be given the form $(s - \alpha_1)(s - \alpha_2)$, $\alpha_{1,2} = (\gamma/2m)(1 \mp \sqrt{1 - 4m/M})$, so that the solution is expressed as

$$\tilde{V}(s) = \frac{2k_B T}{M}\frac{1}{\alpha_2 - \alpha_1}\left(1 + \frac{\gamma}{m}\frac{1}{s}\right)\left(\frac{1}{s-\alpha_2} - \frac{1}{s-\alpha_1}\right). \tag{6}$$

Using the tables of Laplace transforms [12] we get

$$V(t) = \frac{2k_B T}{M}\left\{\frac{\gamma}{m\alpha_1\alpha_2} + \frac{1}{\alpha_2 - \alpha_1}\left[\left(1 + \frac{\gamma}{m\alpha_2}\right)\exp(\alpha_2 t) - \left(1 + \frac{\gamma}{m\alpha_1}\right)\exp(\alpha_1 t)\right]\right\}. \tag{7}$$

The MSD is obtained by simple integration of this equation. It can be used that $\alpha_1\alpha_2 = \gamma^2/(mM)$; it is then seen that the first term in $\{\}$ is the Einstein long-time limit $2k_B T/\gamma$. At short times we return to Eq. (4). If we denote $\mu = \sqrt{1 - 4m/M}$, $D(t)$ can be given a compact form

$$D(t) = \frac{k_B T}{\gamma}\left\{1 - \frac{1}{4\mu}\exp(-\gamma t/2m)\left[(1+\mu)^2\exp(\gamma\mu t/2m) - (1-\mu)^2\exp(-\gamma\mu t/2m)\right]\right\}. \tag{8}$$

If we forget that the motion of the particles was interpreted as induced by particles with smaller mass $m$, an interesting result follows from Eq. (7) in the case when $M < 4m$, i.e. when the roots $\alpha$ are complex. Then the solution describes damped oscillations; e.g. for $D(t)$ we have

$$D(t) = \frac{k_B T}{\gamma}\left\{1 - \exp\left(-\frac{\gamma}{2m}t\right)\times\left[\cos\left(\frac{\gamma t}{2m}\sqrt{\frac{4m}{M}-1}\right) - \frac{2m/M - 1}{\sqrt{4m/M - 1}}\sin\left(\frac{\gamma t}{2m}\sqrt{\frac{4m}{M}-1}\right)\right]\right\}$$

$$\approx \frac{k_B T}{\gamma}\left\{1 + e^{-\gamma t/2m}\left[\sqrt{\frac{m}{M}}\sin\left(\frac{\gamma t}{\sqrt{mM}}\right) - \cos\left(\frac{\gamma t}{\sqrt{mM}}\right)\right]\right\} \text{ (if } M \ll 4m\text{)}. \tag{9}$$

The special case of constant memory kernel $\Gamma$ [13] follows from here when $\gamma/m \to 0$ ($\Gamma = \gamma^2/m = $ const):

$$D(t) \approx \frac{k_B T}{\sqrt{\Gamma M}}\sin\left(\sqrt{\frac{\Gamma}{M}}t\right). \tag{10}$$

## IV. HYDRODYNAMIC BROWNIAN MOTION IN THE MAGNETIC FIELD

The description of the diffusion-like processes with the use of Eq. (1) is not the most general one. In some cases different phenomenological equations can possess a more appropriate description. As discussed in Introduction, the used method is not limited to Eq. (1). Here we shall demonstrate the efficiency of the Vladimirsky approach by the solution of a natural problem that arises in the theory of the BM, namely, the BM with the hydrodynamic memory. This problem can also be formulated using the LE. However, the Stokes friction force $-\gamma v$, which is valid only for the steady motion of the particle (at long times), should be replaced by a force that takes into account the history of the particle motion. This force is a consequence of the inertial effects in the motion of a particle in a liquid, and for incompressible fluids was first derived by Boussinesq [14]. The solution of this problem for BPs was given in the work [15] where, among others, it has been shown that the classic LE gives the correct results only for particles with the density much larger than the density of the surrounding fluid (and simultaneously at short times), or at long times when, however, the Einstein result is valid. Later this task was studied in a number of papers and exactly solved in Ref. [16] (for a review see [17, 18]). With the approach used in the present work the solution can be obtained in a much more simple way than so far. Here we will consider a more complicated problem of the movement of a charged BP in the magnetic field. For the motion along the field or in its absence the previous results [15, 16] will be recovered. The hydrodynamic motion of the BP across the field was already studied in the older paper by Karmeshu [19]. Our solution corrects the results from Ref. [19] in several points.

The Boussinesq force on a spherical particle of radius $R$ is

$$\mathbf{F}(t) = -\gamma \left\{ \mathbf{v}(t) + \frac{\rho R^2}{9\eta} \frac{d\mathbf{v}}{dt} + \sqrt{\frac{\rho R^2}{\pi \eta}} \int_{-\infty}^{t} \frac{d\mathbf{v}}{dt'} \frac{dt'}{\sqrt{t-t'}} \right\}, \tag{11}$$

where the friction factor $\gamma = 6\pi\eta R$, $\rho$ is the density and $\eta$ the viscosity of the solvent. This equation is valid for the times $t \gg R/c$ ($c$ is the sound velocity), i.e., very short times when the compressibility effects play a role are excluded from the consideration. If $Q$ is the charge of the particle of mass $m$ and $\mathbf{B}$ is the constant induction of the magnetic field along the axis $z$, then the LE for the BP is

$$m\dot{\mathbf{v}}(t) + \mathbf{F}(t) = Q\mathbf{v} \times \mathbf{B} + \mathbf{\eta}(t). \tag{12}$$

The projection of this equation onto the axis $z$ does not contain the magnetic force so that along the field we have the motion of a free BP. Equation (2) for $V_z = d\xi_z/dt$ with $\xi_z$ being the MSD in the $z$ direction can be written in the form

$$\dot{V}_z(t) + \frac{1}{\tau}\sqrt{\frac{\tau_R}{\pi}} \int_0^t \frac{\dot{V}_z(t')}{\sqrt{t-t'}} dt' + \frac{1}{\tau} V_z(t) = \frac{2k_B T}{M}, \tag{13}$$

where $M = m + m_s/2$ ($m_s$ is the mass of the solvent displaced by the particle). The characteristic times in this equation are $\tau = M/\gamma$ (the relaxation time of the BP) and $\tau_R = R^2\rho/\eta$ (the vorticity time) The Laplace transformation of Eq. (13) gives

$$\tilde{V}_z(s) = \frac{2k_B T}{M} s^{-1} \left( s + \sqrt{\tau_R} \tau^{-1} s^{1/2} + \tau^{-1} \right)^{-1}. \tag{14}$$

The inverse transform is found after expanding $(..)^{-1}$ in simple fractions $(s - \lambda_{1,2})^{-1}$, where $\lambda_{1,2} = -\left(\sqrt{\tau_R}/2\tau\right)\left(1 \mp \sqrt{1-4\tau/\tau_R}\right)$ are the roots of equation $s + \sqrt{\tau_R}\tau^{-1}s^{1/2} + \tau^{-1} = 0$. Then

$$\tilde{V}_z(s) = \frac{2k_B T}{M} \frac{1}{\lambda_2 - \lambda_1}\left(\frac{1}{\sqrt{s}-\lambda_2} - \frac{1}{\sqrt{s}-\lambda_1}\right), \tag{15}$$

so that [12]

$$V_z(t) = \frac{2k_B T}{M} \frac{1}{\lambda_2 - \lambda_1}\left[\frac{1}{\lambda_2}\exp(\lambda_2^2 t)\mathrm{erfc}\left(-\lambda_2\sqrt{t}\right) - \frac{1}{\lambda_1}\exp(\lambda_1^2 t)\mathrm{erfc}\left(-\lambda_1\sqrt{t}\right)\right]. \tag{16}$$

The VAF is expressed by a similar equation, if we divide $V_z(t)$ by 2 and in [...] replace $1/\lambda_{1,2}$ with $\lambda_{1,2}$. This expression exactly corresponds to the solutions found in Refs. [15, 16] by different methods but differs from the solution [19] due to the difference in the roots $\lambda_{1,2}$. The solution at long times contains the so called long-time tails that became famous after their discovery in the computer experiments [20, 21]. For the VAF at $t \to \infty$ it follows from Eq. (16) that the longest-lived tail is $\sim t^{-3/2}$,

$$\Phi(t) \approx \frac{k_B T}{2\sqrt{\pi}M}\frac{\tau\sqrt{\tau_R}}{t^{3/2}}\left[1 - \frac{3}{2}\left(1 - 2\frac{\tau}{\tau_R}\right)\frac{\tau_R}{t} + ...\right]. \tag{17}$$

The MSD that is obtained by integrating Eq. (16) corresponds very well to experiment [22].

For the particle motion across the magnetic field we have from Eq. (12) two equations for the $x$ and $y$ components of the velocity and after the application of Vladimirsky rule two equations for the quantities $V_x = d\xi_x/dt$ and $V_y = d\xi_y/dt$ that determine the MSD in the $x$ and $y$ directions. The sum $\xi_{xy} = \xi_x + \xi_y$ gives the MSD of the particle across the field. In the Laplace transformation we obtain for $\tilde{V}_{xy}(s) = \mathcal{L}\{V_x(t) + V_y(t)\}$

$$\tilde{V}_{xy}(s) = \frac{4k_B T}{Ms}\frac{\psi(s)}{\psi^2(s) + \tau_c^{-2}}, \tag{18}$$

where $\psi(s) = s + \tau^{-1}\sqrt{\tau_R s} + \tau^{-1}$ and the new characteristic time $\tau_c$ is connected to the cyclotron frequency $\omega_c = QB/M = 1/\tau_c$. The roots $\lambda_{1,2}$ of $\psi(s)$ have been determined after Eq. (14) and for the four roots of the denominator $\psi^2(s) + \tau_c^{-2}$ we have $2\kappa_{1,2} = \lambda_1 + \lambda_2 \pm \sqrt{(\lambda_1 - \lambda_2)^2 - 4i\omega_c}$ and $2\kappa_{3,4} = \lambda_1 + \lambda_2 \pm \sqrt{(\lambda_1 - \lambda_2)^2 + 4i\omega_c}$. Using them, the $s$-dependent fraction in Eq. (18) can be decomposed in simple fractions,

$$\frac{2\psi(s)}{\psi(s)^2 + \tau_c^{-2}} = \frac{1}{\kappa_2 - \kappa_1}\left(\frac{1}{\sqrt{s}-\kappa_2} - \frac{1}{\sqrt{s}-\kappa_1}\right) + \frac{1}{\kappa_4 - \kappa_3}\left(\frac{1}{\sqrt{s}-\kappa_4} - \frac{1}{\sqrt{s}-\kappa_3}\right),$$

after which the Laplace transform tables can be used [12]. This yields

$$V_{xy}(t) = \frac{2k_B T}{M}\left\{\left(\frac{1}{\kappa_1\kappa_2} + \frac{1}{\kappa_2 - \kappa_1}\left[f(\kappa_2, t) - f(\kappa_1, t)\right]\right) + (\kappa_1 \to \kappa_3, \kappa_2 \to \kappa_4)\right\}, \tag{19}$$

where $f(\kappa,t) = \kappa^{-1}\exp(\kappa^2 t)\mathrm{erfc}(-\kappa\sqrt{t})$. The MSD is readily obtained by the integration of $V_{xy}(t)$ from 0 to $t$. Using

$$\varphi(\kappa,t) = \int_0^t f(\kappa,t')\mathrm{d}t' = \frac{1}{\kappa^3}\left[\exp\left(\kappa^2 t\right)\mathrm{erfc}(-\kappa\sqrt{t})-1\right] - \frac{2}{\kappa^2}\sqrt{\frac{t}{\pi}}$$

one finds

$$\begin{aligned}\xi_{xy}(t) = \frac{2k_B T}{M}\Bigg\{&\left(\frac{t}{\kappa_1\kappa_2} + 2\sqrt{\frac{t}{\pi}}\frac{\kappa_1+\kappa_2}{(\kappa_1\kappa_2)^2} + \frac{\kappa_1^2+\kappa_1\kappa_2+\kappa_2^2}{(\kappa_1\kappa_2)^3}\right.\\&+ \frac{1}{\kappa_2-\kappa_1}\left[\frac{1}{\kappa_2^3}\exp\left(\kappa_2^2 t\right)\mathrm{erfc}\left(-\kappa_2\sqrt{t}\right) - \frac{1}{\kappa_1^3}\exp\left(\kappa_1^2 t\right)\mathrm{erfc}\left(-\kappa_1\sqrt{t}\right)\right]\bigg)\\&+ \ (\kappa_1 \to \kappa_3, \kappa_2 \to \kappa_4)\Bigg\}.\end{aligned} \qquad (20)$$

In the absence of the field ($\omega_c = 0$) $\kappa_1 = \kappa_3 = \lambda_1$, $\kappa_2 = \kappa_4 = \lambda_2$, and $\xi_x = \xi_y = \xi_z$, with $\xi_z$ from Eq. (14). Then for long times, up to the longest-lived tail,

$$\xi_{xy}(t) \approx \frac{4k_B T}{M}\frac{\tau t}{1+(\tau/\tau_c)^2}\left(1 - 2\sqrt{\frac{\tau_R}{\pi t}}\frac{1-(\tau/\tau_c)^2}{1+(\tau/\tau_c)^2} + ...\right), \quad t \to \infty. \qquad (21)$$

Even the Einstein limit $\sim t$ differs from the previous result [19] (the agreement is only when $M = m$ (or at $\omega_c = 0$)). At short times we have the same result as in Eq. (4) (independent of the magnetic field) but with a different $M$, which is now $M = m + m_s/2$. This apparent contradiction with the equipartition theorem is only a consequence of the assumption of solvent incompressibility, due to which the limit $t \to 0$ cannot be accomplished. The correct result that contains only the mass of the particle, $m$, is achieved when the compressibility of the solvent is taken into account (on the time scale $\sim 10^{-12}$ s characterizing the collisions of the BP with the surrounding molecules the VAF decays from the equipartition value $k_B T/m$ to $k_B T/(m + m_s/2)$ through the emission of sound waves). This is one more task that is straightforwardly solvable by the present method. Recently [23], the motion of a charged BP in a magnetic field was studied in the case when the aftereffect exponentially decreases in the time. Also this problem can be more effectively solved by the proposed method.

## V. CONCLUSION

In conclusion, we have described an effective approach to the solution of the problems of anomalous Brownian motion modeled using the generalized Langevin equation. The method of its solution consists in converting this stochastic integro-differential equation into a deterministic one. It allowed us to obtain some general properties of the anomalous diffusion in a very simple way and to solve the specific problem when the Brownian particle is driven by an exponentially correlated stochastic force. The description of diffusion-like processes with the use of the GLE is by no means the most general one. In many cases different phenomenological equations could possess a more appropriate model. It is thus important that the used method is not limited to the equations of GLE type. We have demonstrated it on the problem of the hydrodynamic Brownian motion of charged particles in magnetic field. Other applications that are currently of great interest concern, for example, the anomalous motion of colloidal particles under influence of various external fields [24, 25], the motion of Brownian particles dragged by optical tweezers [26, 27], the behavior of mesoscale electric circuits in contact with the thermal bath [28, 29], the motion of magnetic domain walls [30], or the dynamics of polymers [18, 31]. In our opinion, future effort should be first of all oriented on the generalization of the proposed method to situations when quantum effects are significant [28], and to overcome its restriction to linear models.

## ACKNOWLEDGMENTS

This work was supported by the Agency for the Structural Funds of the EU within the projects NFP 26220120021 and 26220120033, and by the grant VEGA 1/0300/09.

___________________________________________

[1] R. Klages, G. Radons, and I. M. Sokolov (eds.), Anomalous Transport: Foundations and Applications (Wiley-VCH, Berlin, 2008).
[2] A. Einstein, Ann. Phys. **17**, 549 (1905).
[3] R. Kubo, Rep. Prog. Phys. **29**, 255 (1966).
[4] W. T. Coffey, Yu. P. Kalmykov, and J. T. Waldron, The Langevin Equation (World Scientific, New Jersey, 2005).
[5] P. Langevin, C. R. Acad. Sci. (Paris) **146**, 530 (1908).
[6] V. V. Vladimirsky, Zhur. Ekper. Teor. Fiz. **12**, 199 (1942).
[7] V. Balakrishnan, Pramana **12**, 301 (1979).
[8] P. P. J. M. Schram and I. P. Yakimenko, Physica A **260**, 73 (1998).
[9] R. Morgado, F. A. Oliveira, G. G. Batrouni, and A. Hansen, Phys. Rev. Lett. **89**, 100601 (2002).
[10] L. C. Lapas, R. Morgado, M. H. Vainstein, J. M. Rubí, and F. A Oliveira, Phys. Rev. Lett. **101**, 230602 (2008).
[11] E. Lutz, Phys. Rev. E **64**, 051106. 2001.
[12] M. Abramowitz and I. A. Stegun, Handbook of Mathematical Functions (National Bureau of Standards, Washington, DC, 1964).
[13] I. Goychuk, Phys. Rev. E **80**, 046125 (2009).
[14] J. Boussinesq, C. R. Acad. Sci. Paris **100**, 935 (1885).
[15] V. Vladimirsky and Ya. Terletzky, Zhur. Eksp. Teor. Fiz. **15**, 259 (1945).
[16] E. J. Hinch, J. Fluid. Mech. **72**, 499 (1975).
[17] V. Lisy and J. Tothova, arXiv:cond-mat/0410222v1 [cond-mat.stat-mech].
[18] J. Tothova, V. Lisy, and A.V. Zatovsky, J. Chem. Phys. **119**, 13135 (2003); V. Lisy, J. Tothova, and A.V. Zatovsky, J. Chem. Phys. **121**, 10699 (2004).
[19] Karmeshu, J. Phys. Soc. Jpn. **34**, 1467 (1973).
[20] A. Rahman, Phys. Rev. A **136**, 405 (1964).
[21] B. J. Alder and T. E. Wainwright, Phys. Rev. Lett. **18**, 988 (1967).
[22] B. Lukić, S. Jeney, C. Tischer, A. J. Kulik, L. Forró, E. L. Florin, Phys. Rev. Lett. **95**, 160601 (2005).
[23] F. N. C. Paraan, M. P. Solon, and J. P. Esguerra, Phys. Rev. E **77**, 022101 (2008).
[24] D. M. Carberry, J. C. Reid, G. M. Wang, E. M. Sevick, D. J. Searles, and D. J. Evans, Phys. Rev. Lett. **92**, 140601 (2004).
[25] H. Löwen, J. Phys.: Condens. Matter **21**, 474203 (2009).
[26] R. Van Zon and E. G. D. Cohen, Phys. Rev. E **67**, 046102 (2003).
[27] T. Li, S. Kheifets, D. Medellin, and M. G. Raizen, www.sciencexpress.org/20 May 2010 /10.1126/ science.1189403.
[28] A. E. Allahverdyan, Th. M. Nieuwenhuizen, Phys. Rev. B **66**, 115309 (2002).
[29] N. Garnier and S. Ciliberto, Phys. Rev. E **71**, 060101(R) (2005).
[30] E. Saitoh, H. Miyajima, T. Yamaoka, and G. Tatara, Nature **432**, 203 (2004).
[31] D. Panja, arXiv:1004.0935v3 [cond-mat.soft], submitted to JSTAT.